\documentclass[reprint,
 amsmath,amssymb,
prb,
]{revtex4-1}

\usepackage{graphicx}
\usepackage{dcolumn}
\usepackage{longtable}
\usepackage{lipsum}
\usepackage{color}
\begin{document}


\title{ Josephson effect in two-band superconductors}
\author{Akihiro Sasaki$^{1}$}
\author{Satoshi Ikegaya$^{2}$}
\author{Tetsuro Habe$^{1}$}
\author{Alexander A. Golubov$^{3,4}$}
\author{Yasuhiro Asano$^{1,5}$}
\affiliation{$^{1}$ Department of Applied Physics,
Hokkaido University, Sapporo 060-8628, Japan\\
$^{2}$ Max-Planck-Institut f\"{u}r Festk\"{o}rperforschung,
Heisenbergstrasse 1, D-70569 Stuttgart, Germany\\
$^{3}$Moscow Institute of Physics and Technology, 141700 Dolgoprudny, Russia\\
$^{4}$Faculty of Science and Technology and MESA+ Institute for Nanotechnology,
University of Twente,
7500 AE Enschede, The Netherlands\\
$^{5}$Center of Topological Science \& Technology,
Hokkaido University, Sapporo 060-8628, Japan\\
}%

\date{\today}

\begin{abstract}
We study theoretically the Josephson effect between two time-reversal two-band superconductors,
where we assume the equal-time spin-singlet $s$-wave pair potential in each conduction band.
The superconducting phase at the first band $\varphi_1$ and that at the second band $\varphi_2$
characterize a two-band superconducting state.
We consider a Josephson junction where an insulating barrier separates two such two-band
superconductors.
By applying the tunnel Hamiltonian description, the Josephson current is calculated
in terms of the anomalous Green's function on either side of the junction.
We find that the Josephson current consists of three components
 which depend on three types of phase differences across the
 junction: the phase difference at the first band $\delta\varphi_1$,
the phase difference at the second band $\delta\varphi_2$, and the difference
at the center-of-mass phase $\delta(\varphi_1+\varphi_2)/2$.
A Cooper pairs generated by the band hybridization carries the last current component.
In some cases, the current-phase relationship deviates from the sinusoidal function
as a result of time-reversal symmetry breaking down.
\end{abstract}


\maketitle

\section{Introduction}

The Josephson effect is a fundamental property of all superconducting junctions
consisting of more than one superconductor~\cite{josephson:physlett1962}.
The dissipationless electric current ($J$) flows through the junction in the presence of
difference in superconducting phases across the junction $\delta\varphi$.
The current-phase relationship (CPR) depends sensitively on electric and magnetic properties
of a material sandwiched by two superconductors~\cite{likharev:rmp1979,golubov:rmp2004}.
When the material is an insulator, it is well established that CPR is sinusoidal, (i.e.,
$J=J_0 \sin\delta\varphi$)~\cite{ambegaokar:prl1963}. The relationship suggests
that the junction energy is its minimum at $\delta\varphi=0$.

The expression of CPR is expected to be complicated
when the junction energy depends on more than two superconducting phases.
For instance, a superconducting state is characterized by two phases
in a two-band superconductor such as MgB$_2$~\cite{mgb2:akimitsu2001,mgb2:louie2002}
and iron pnictides~\cite{pnictide:hosono2008,hosono:physicac2015}.
Namely the phase of the pair potential in the first band $\varphi_1$ and
that of the second band $\varphi_2$ characterize a two-band superconductivity.
Theories~\cite{kuroki:prl2008,raghu:prb2008,kontani:prl2010} have suggested
that either an $s_{++}$ state of $\varphi_1=\varphi_2$ or an $s_{+-}$ state
of $\varphi_1=\varphi_2+\pi$ can be realized in pnictides.
When such a two-band superconductor couples to a single-band $s$-wave superconductor
 through an insulator, the junction energy depends on the phase of the single-band
 superconductor $\varphi_s$, $\varphi_1$ and $\varphi_2$.
Actually, theoretical studies on such Josephson junctions have shown the complicated
energy diagrams as a function of
these phases~\cite{ng:epl2009,stanev:prb2010,sperstad:prb2009,yerin:ltp2017}.
Especially, an $s_{+-}$ state frustrates the junction energy and stabilizes
a spontaneously time-reversal
symmetry breaking state near the junction interface~\cite{ng:epl2009,stanev:prb2010}.
More complicated CPR would be expected in a Josephson junction consisting of two pnictide
superconductors\cite{nappi:PRB2015} because four superconducting phases enter
the junction energy.
At present, this is still an open issue.

In this paper, we discuss the Josephson current between two time-reversal
two-band (two-orbital) superconductors.
We assume an equal-time spin-singlet $s$-wave order parameter
in each conduction band and consider the band hybridization as well as the band asymmetry.
The coupling between the two superconductors is
described by the tunneling Hamiltonian which includes the band-diagonal hopping term.
 The Josephson current is calculated by using the anomalous Green's function
 on either side of the junction.
 In the absence of band hybridization, the Josephson current consists of
 two conventional components:
 $J_1\sin(\delta\varphi_1)$ and $J_2\sin(\delta\varphi_2)$, where $\delta\varphi_\lambda$
is the phase difference of the pair potential at the $\lambda$-th band for $\lambda=1-2$.
The band hybridization generates the \textsl{interband} pairing correlations
which carry the Josephson current of $J_{12}\sin[(\delta\varphi_1+\delta\varphi_2)/2]$.
We will show that this additional component exhibits unusual property and can be a
source of time-reversal symmetry breaking of the junction.
 We also discuss a role of
an odd-frequency Cooper pair~\cite{berezinskii:jetplett1974} in the Josephson effect
of two-band superconductors~\cite{BSchaffer:prb2013,asano:prb2015,asano:prb2018}.

This paper is organized as follows. In Sec.~II, we describe
a time-reversal superconducting state in a two-band
superconductor in terms of a microscopic Hamiltonian.
The solution of the Gor'kov equation is also presented.
In Sec.~III, we formulate the Josephson current by using tunnel Hamiltonian
between two superconductors. On the basis of the analytical results, we discuss
curent-phase relationship of the Josephson current.
The dependence of the Josephson current on temperature is shown by solving
the gap equation numerically.
In Sec.~V, we discuss the junction energy phenomenologically.
The conclusion is given in Sec.~V.
Throughout this paper, we use the units of $k_\mathrm{B}=c=\hbar=1$, where $k_{\mathrm{B}}$
is the Boltzmann constant and $c$ is the speed of light.

\section{Two-band superconductor}\label{model}

\subsection{Hamiltonian}

The BCS Hamiltonian for a two-band superconductor is described by,
\begin{align}
&\mathcal{H} = \int d\boldsymbol{r}
\Psi^\dagger(\boldsymbol{r})\, H \, \Psi(\boldsymbol{r}),\\
&\Psi(\boldsymbol{r})=
\left[
 \psi_{1,\uparrow}(\boldsymbol{r}),
 \psi_{2,\uparrow}(\boldsymbol{r}),
 \psi_{1,\downarrow}^\dagger(\boldsymbol{r}),
 \psi_{2,\downarrow}^\dagger(\boldsymbol{r})
 \right]^{\mathrm{T}},\label{h_s}\\
& H
= \left[ \begin{array}{cccc}
 \xi_1(\boldsymbol{r})
  & ve^{i\theta}  & |{\Delta}_1| e^{i\varphi_1}  & 0\\
 ve^{-i\theta}  & \xi_2(\boldsymbol{r}) & 0 &|{\Delta}_2| e^{i\varphi_2} \\
|{\Delta}_1| e^{-i\varphi_1}  & 0 & -\xi_1(\boldsymbol{r})  &
 -ve^{-i\theta} \\
0&  |{\Delta}_2| e^{-i\varphi_2}  & -ve^{i\theta}  & -\xi_2(\boldsymbol{r})
\end{array}\right], \label{hbdg_real} \\
&\xi_1(\boldsymbol{r}) = - \frac{\nabla^2}{2m} -\gamma - \epsilon_F, \;
\xi_2(\boldsymbol{r}) = - \frac{\nabla^2}{2m} +\gamma - \epsilon_F,
\end{align}
where $\psi_{\lambda,\sigma}^\dagger(\boldsymbol{r})$ ($\psi_{\lambda,\sigma}(\boldsymbol{r})$) is the
creation (annihilation) operator of an electron with spin $\sigma$ ($=\uparrow$ or $\downarrow$) at the $\lambda$ th conduction band, $ve^{i\theta}$ denotes the hybridization between the two bands,
and $\mathrm{T}$ means the transpose of a matrix.
The kinetic energy of an electron at the $\lambda$ th band is represented by
 $\xi_\lambda(\boldsymbol{r})$, where $\gamma$ describes the asymmetry in the two band.
We assume a uniform spin-singlet $s$-wave pair potential for each conduction band which is defined by
\begin{align}
\Delta_\lambda = g_\lambda \left\langle \psi_{\lambda, \uparrow}(\boldsymbol{r})
\psi_{\lambda, \downarrow}(\boldsymbol{r}) \right\rangle, \label{delta_def}
\end{align}
where $g_\lambda>0$ represents the attractive interaction between two electrons
on the fermi surface of the $\lambda$ th band.

In this paper, we do not consider any spin-dependent potentials.
Therefore, the Hamiltonian In Eq.~(\ref{hbdg_real}) is represented in $4 \times 4$ matrix
form by extracting
spin $\uparrow$ sector from the particle space and spin $\downarrow$ sector from the hole space.
Time-reversal symmetry of such a Bogoliubov-de Gennes Hamiltonian ${H}_A$
 is represented by
\begin{align}
\mathcal{T}\, H_A\, \mathcal{T}^{-1} =H_A,\quad
{\mathcal{T}} = \hat{\rho}_0\, \hat{\tau}_0\, \mathcal{K}, \label{trs}
\end{align}
where $\mathcal{K}$ means the complex conjugation,
 $\hat{\rho}_0$ and $\hat{\tau}_0$ are the unit matrices in two-band space and
 particle-hole space, respectively.
Generally speaking, the two superconducting phases in the Hamiltonian
can be removed by a gauge transformation
$H_A = U H_A^\prime U^\ast$, where $U$ is a unitary matrix satisfying $U^\ast=U^{-1}$.
When all the elements of $H_A^\prime$ are real numbers,
$H_A$ preserves time-reversal symmetry because unitary transformation does not
 change physics.
Actually, it is possible to show the relation
\begin{align}
\mathcal{T}_U\, H_A\, \mathcal{T}_U^{-1} =H_A,\quad
\mathcal{T}_U = U^2\, \mathcal{K},
\end{align}
where $\mathcal{T}_U$ is a gauge-dependent time-reversal operator.
At the Hamiltonian $H$ in Eq.~(\ref{hbdg_real}),
a unitary operator
\begin{align}
{U} &= \textrm{diag}[e^{i\varphi_1/2}, e^{i\varphi_2/2},
e^{-i\varphi_1/2}, e^{-i\varphi_2/2}],
\end{align}
removes two superconducting phases.
Under such gauge transformation, however, the phase of the hybridization
is transformed as
\begin{align}
\theta \to \theta - \frac{\varphi_1}{2} + \frac{\varphi_2}{2}.
\end{align}
Therefore, we conclude that Eq.~(\ref{hbdg_real}) preserves time-reversal symmetry only when
\begin{align}
2\theta - \varphi_1 + \varphi_2=2\pi n, \label{trs_cond}
\end{align}
is satisfied~\cite{asano:prb2018}.
The phases of the two pair potentials and that of the hybridization are linked to
one another when the superconductor preserves time-reversal symmetry.

Even in the absence of hybridization, the standard mean-field theory
in two-band system contains interaction terms between $\Delta_1$
and $\Delta_2$ ~\cite{leggett:PTP1966} as,
\begin{align}
&{g_{12}^\ast}
\left\langle
 \psi_{1, \uparrow}(\boldsymbol{r})
\psi_{1, \downarrow}(\boldsymbol{r})
\right\rangle  \psi_{2, \downarrow}^\dagger(\boldsymbol{r})
\psi_{2, \uparrow}^\dagger(\boldsymbol{r})\nonumber\\
&
+
{g_{12}}\left\langle
 \psi_{2, \uparrow}(\boldsymbol{r})
\psi_{2, \downarrow}(\boldsymbol{r})
\right\rangle  \psi_{1, \downarrow}^\dagger(\boldsymbol{r})
\psi_{1, \uparrow}^\dagger(\boldsymbol{r}) +\textrm{H. c.},
\end{align}
where the phase of the interaction potential $g_{12}=|g_{12}|e^{2i\theta}$
relates $\varphi_1$ and $\varphi_2$~\cite{asano:prb2018}.
In this paper, we do not consider these interaction terms because
they do not affect the main conclusion.

\subsection{Gor'kov equation}\label{gorkov}
The Gor'kov equation in the Matsubara representation reads,
\begin{align}
\left[ i\omega_n - {H}(\boldsymbol{k}) \right]
\left[ \begin{array}{cc} \hat{\mathcal{G}}& \hat{\mathcal{F}} \\
\underline{\hat{\mathcal{F}} } & -\underline{\hat{\mathcal{G}}}
\end{array}\right]_{(\boldsymbol{k}, i\omega_n)} = \check{1},
\end{align}
where $\omega_n=(2n+1) \pi T$ is the fermionic Matsubara frequency with $T$ being a temperature.
The particle-hole symmetry implies the relations,
\begin{align}
\underline{\hat{\mathcal{G}}}(\boldsymbol{k}, i\omega_n) =&
\hat{\mathcal{G}}^\ast(-\boldsymbol{k}, i\omega_n),
\\
\underline{\hat{\mathcal{F}}}(\boldsymbol{k}, i\omega_n) =&
\hat{\mathcal{F}}^\ast(-\boldsymbol{k}, i\omega_n),
\end{align}
where $\hat{\cdots}$ denotes $ 2 \times 2$ matrix in two-band space.
The anomalous Green function is represented as
\begin{align}
&\hat{\mathcal{F}}(\boldsymbol{k},\omega_n) = \frac{1}{Z}\left[\begin{array}{cc}
f_{1}  & f_{\mathrm{e}}-if_{\mathrm{o}} \\
f_{\mathrm{e}}+if_{\mathrm{o}}  & f_{2}
\end{array}\right],\label{f-function}\\
&Z= Z_1\, Z_2  + 2\, v^2 \, ( \omega_n^2 -  \xi_1\, \xi_2 + |\Delta_1|\,|\Delta_2|) +v^4, \\
&Z_\lambda= \xi_\lambda^2+|\Delta_\lambda|^2 + \omega_n^2,
\end{align}
\begin{align}
f_1(\boldsymbol{k},\omega_n)=& -\left[ Z_2|\Delta_1| +v^2|\Delta_2| e^{i2 n \pi}
\right]\, e^{i\varphi_1},\\
f_2(\boldsymbol{k},\omega_n)=& -\left[ Z_1|\Delta_2| +v^2|\Delta_1|e^{-i2 n \pi} \right]\, e^{i\varphi_2},\\
f_{\mathrm{e}}(\boldsymbol{k},\omega_n)=& \,v \, [\xi_1|\Delta_2| +\xi_2|\Delta_1|]\; e^{i(\varphi_1+\varphi_2)/2} \; e^{i n \pi},\\
f_{\mathrm{o}}(\boldsymbol{k},\omega_n)=& \, \omega_n\, v\, [|\Delta_1| -|\Delta_2|]\; e^{i(\varphi_1+\varphi_2)/2}
\; e^{i n \pi},
\end{align}
where $n\pi$ is derived from the right hand side of Eq.~(\ref{trs_cond}).
The dispersion in the two bands are described by
$\xi_1=\xi_{\boldsymbol{k}}-\gamma$ and $\xi_2=\xi_{\boldsymbol{k}}+\gamma$ with
 $\xi_{\boldsymbol{k}}=\boldsymbol{k}^2/2m -\epsilon_F$.
Here we also supply the expression of the normal Green function
\begin{align}
\hat{\mathcal{G}}(\boldsymbol{k},\omega_n) =& \frac{1}{Z}\left[\begin{array}{cc}
g_{1}  & g_{12} \\
g_{21}  & g_{2}
\end{array}\right],\label{gnormal}\\
g_1(\boldsymbol{k},\omega_n)=& -\left[ (i\omega_n+\xi_1) Z_2 + (i\omega_n -\xi_2) v^2 \right],\\
g_2(\boldsymbol{k},\omega_n)=& -\left[ (i\omega_n+\xi_2) Z_1 + (i\omega_n -\xi_1) v^2 \right],\\
g_{12}(\boldsymbol{k},\omega_n)=& A\, v\, e^{i\theta},\\
g_{21}(\boldsymbol{k},\omega_n)=& A\, v\, e^{-i\theta},\\
%
A=&(i\omega+\xi_1)(i\omega+\xi_2) - v^2 - |\Delta_1||\Delta_2|.
\end{align}

The gap equations are represented as
\begin{align}
\Delta_\lambda=& -g_\lambda T\sum_{\omega_n} \frac{1}{V_{\mathrm{vol}}}\sum_{\boldsymbol{k}} f_\lambda(\boldsymbol{k}, \omega_n)
\end{align}
and result in
\begin{align}
\Delta_1=&T\sum_{\omega_n} \frac{1}{V_{\mathrm{vol}}}\sum_{\boldsymbol{k}} \frac{g_1}{Z}
\left[ Z_2|\Delta_1| + v^2|\Delta_2|\right]e^{i\varphi_1}, \label{gap1}\\
\Delta_2=&  T\sum_{\omega_n} \frac{1}{V_{\mathrm{vol}}}\sum_{\boldsymbol{k}} \frac{g_2}{Z}
\left[ Z_1|\Delta_2| + v^2|\Delta_1|\right]e^{i\varphi_2}.\label{gap2}
\end{align}
Under the condition in Eq.~(\ref{trs_cond}), these equations have stable solutions
 of $|\Delta_1|$ and $|\Delta_2|$.

\section{Josephson Effect}
\subsection{Current formula}
We discuss the Josephson effect between two
superconductors, where the superconducting states of each
superconductor is described by Eq.~(\ref{hbdg_real}).
We assume that $\xi_1$, $\xi_2$,
$|\Delta_1|$, $|\Delta_2|$ and $v$ are common in the two superconductors.
The two superconductors are connected by the Hamiltonian
\begin{align}
\mathcal{H}_{T} =& \sum_{\boldsymbol{k}, \boldsymbol{p}, \sigma}
\left[ \psi_{L, 1, \boldsymbol{p}, \sigma}^\dagger,
 \psi_{L, 2, \boldsymbol{p}, \sigma}^\dagger
\right] \, \hat{t}_\mathrm{T}\,
\left[ \begin{array}{c}
 \psi_{R, 1, \boldsymbol{k}, \sigma} \\ \psi_{R, 2, \boldsymbol{k}, \sigma}
\end{array}\right] \nonumber\\
&+ \mathrm{H. c.},\\
\hat{t}_\mathrm{T}=& \left[ \begin{array}{cc}
 t_1  & 0  \\ 0  & t_2
\end{array}\right],
\end{align}
where $L$ ($R$) indicates the superconductor on the left (right) hand side of
an insulating barrier, $t_1$ and $t_2$ denote the tunneling elements through
the barrier at the first band and that
at the second band, respectively.
We assume that $t_1$ and $t_2$ are real number and are independent of
the wave numbers of an electron.
The total Hamiltonian reads,
\begin{align}
\mathcal{H}_{\mathrm{JJ}} = \mathcal{H}_L + \mathcal{H}_R + \mathcal{H}_T, \label{h_jj}
\end{align}
where $\mathcal{H}_{L(R)}$ is given by Eq.~(\ref{h_s})
with $\varphi_1 \to \varphi_{1 L(1R)}$, $\varphi_2 \to \varphi_{2L(2R)}$, and
$\theta_1 \to \theta_{L(R)}$.
The Josephson current is represented by~\cite{mahan}
\begin{align}
J=&e\, \mathrm{Im}
\sum_{\boldsymbol{k},\boldsymbol{p}, \sigma}
 T\sum_{\omega_n} \mathrm{Tr} \left[
\hat{t}_\mathrm{T}\,
\hat{\underline{\mathcal{F}}}_R(\boldsymbol{k}, i\omega_n)
\,
\hat{t}_\mathrm{T}
\,
\hat{{\mathcal{F}}}_L(\boldsymbol{p}, i\omega_n)
 \right], \label{formula}
\end{align}
where $\mathrm{Tr}$ means the trace over
two-band degree of freedom.
Since time-reversal symmetry is preserved in each superconductor, the phases satisfy
\begin{align}
&2\theta_L-\varphi_{1L}+\varphi_{2L}=2\pi n_L,\label{trs_l}\\
& 2\theta_R-\varphi_{1R}+\varphi_{2R}=2\pi n_R. \label{trs_r}
\end{align}

By substituting the anomalous Green's function in Eq.~(\ref{f-function})
into the formula in Eq.~(\ref{formula}), we obtain
\begin{align}
J
=&J_1 \sin(\varphi_{1L} -\varphi_{1R})
+ J_2 \sin(\varphi_{2L}-\varphi_{2R})\nonumber\\
&+J_{12} \sin\left[\bar{\varphi}_L -\bar{\varphi}_R+(n_L-n_R)\pi \right],\\
\bar{\varphi}_L=&\frac{\varphi_{1L}+\varphi_{2L}}{2}, \;
\bar{\varphi}_R=\frac{\varphi_{1R}+\varphi_{2R}}{2},
\end{align}
where the amplitudes are given by
\begin{align}
J_1=&J_0 \frac{T}{T_c} \sum_{\omega_n} \frac{t_1^2}{t_1^2+t_2^2}
\left[
\sum_{\boldsymbol{k}} \frac{ Z_2|\Delta_1| + v^2|\Delta_2| }{\pi N_0 Z}
\right]^2,\label{j1}\\
J_2=&J_0 \frac{T}{T_c} \sum_{\omega_n} \frac{t_2^2}{t_1^2+t_2^2}
\left[
\sum_{\boldsymbol{k}} \frac{ Z_1|\Delta_2| + v^2|\Delta_1| }{\pi N_0 Z}
\right]^2,\\
J_{12}=&J_0 \frac{T}{T_c} \sum_{\omega_n} \frac{2t_1 t_2}{t_1^2+t_2^2} v^2
 \left[
\left\{
\sum_{\boldsymbol{k}} \frac{ \xi_2|\Delta_1| + \xi_1|\Delta_2| }{\pi N_0 Z}
\right\}^2  \right.\nonumber\\
&\left.-\left\{
\sum_{\boldsymbol{k}} \frac{ \omega_n(|\Delta_1| -|\Delta_2|) }{\pi N_0 Z}
\right\}^2
\right],\label{j12}\\
J_0=&\frac{\pi T_c}{2e R_N}, \quad
G_{\mathrm{N}}= {4\pi e^2}(t_1^2+t_2^2) N_0^2=R^{-1}_{\mathrm{N}}.
\end{align}
The normal conductance $G_{\mathrm{N}}$ is calculated from the normal Green function
in Eq.~(\ref{gnormal}) at $\Delta_1=\Delta_2=0$ and $N_0$ is the normal density of
states per spin at the fermi level.
The first term proportional to $J_1$ is a result of the
tunneling of a Cooper pair between the first
conduction band of the two superconductors.
In a similar way, the second term represents the Josephson current between the second bands.
The third term proportional to $J_{12}$ is the tunneling of an
induced Cooper pair by the band hybridization.
The hybridization generates two pairing correlations: $f_e$ and $f_o$ in Eq.~(\ref{f-function}).
The component $f_e$ belongs to even-frequency symmetry class as shown in Eq.~(\ref{f-function})
and gives the first term in Eq.~(\ref{j12}).
On the other hand, the component $f_o$ belongs to odd-frequency symmetry class
as show in Eq.~(\ref{f-function})~\cite{asano:prb2015} and gives the second term in Eq.~(\ref{j12}).
As a result, $J_{12}$ can be negative when the amplitude of $f_o$ is larger than that of $f_e$.
It has been known that an odd-frequency Cooper pair indicates the paramagnetic response
to magnetic field~\cite{asano:prl2011,asano:prb2015,suzuki:prb2014} and favors the
spatial gradient in the phase.

%
%
%
%

\subsection{Dependence of $J$ on temperature}
\begin{figure}[tbh]
\begin{center}
\includegraphics[width=8.5cm]{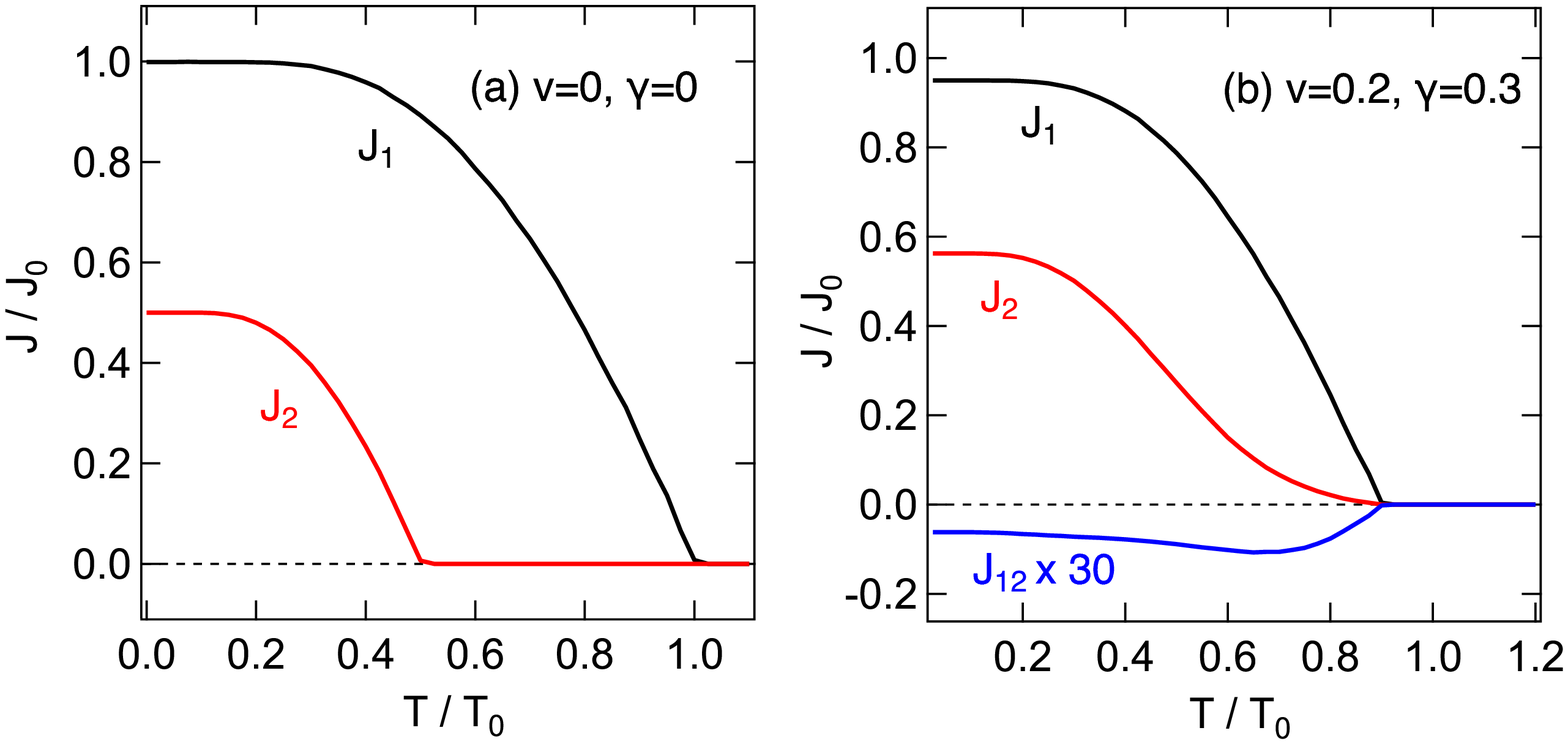}
\end{center}
\caption{
The dependence of $J_1$, $J_2$ and $J_{12}$ on temperature for
$g_2=0.5 g_1$ and $t_1=t_2=\epsilon_F$.
The parameters are fixed at
$\gamma=v=0$ in (a), and $\gamma=0.2 \epsilon_F$ and $v=0.3 \epsilon_F$ in (b).
 }
\label{fig:tdepj}
\end{figure}

The amplitudes of the current in Eqs.~(\ref{j1})-(\ref{j12}) are plotted as a function of temperature in Fig.~\ref{fig:tdepj},
where we choose $g_2=0.5 g_1$ and $t_1=t_2$.
The pair potentials $\Delta_1$ and $\Delta_2$ are calculated by solving
the gap equations in Eqs.~(\ref{gap1}) and (\ref{gap2}).
The transition temperature at the first band for $v=\gamma=0$ is indicated by $T_0$.
In Fig.~\ref{fig:tdepj}(a), $J_1$ and $J_2$ are shown for $v=\gamma=0$.
The two current component are independent of each other and $J_{12}=0$
in the absence of hybridization. The transition temperature at the first band is
indicated by $T_0$.
 The transition temperature at the second band
is 0.5$T_0$ as a result of $g_2=0.5 g_1$.
%
%
In Fig.~\ref{fig:tdepj}(b), we show the results of $J_1$, $J_2$ and $J_{12}$
for $v/2\pi T_0 =0.2$ and $\gamma/2\pi T_0 =0.3$.
The transition temperature in this case is about 0.9$T_0$. The all current
components are finite at the temperature below 0.9$T_0$.
The results of $J_{12}$ are amplified by 30 for better visibility and are negative
at this parameter choice.
\subsection{Current-phase relationship}

Three phases $\varphi_{1j}$, $\varphi_{2j}$, and $\theta_{j}$
characterize a two-band superconducting state for $j=L$ and $R$.
The condition derived from time-reversal symmetry in Eqs.~(\ref{trs_l}) and (\ref{trs_r})
reduces the degree of freedom in phases to two.
The phase difference $\delta\theta=\theta_L-\theta_R$ is a parameter characterizing a junction
because $\theta_{L(R)}$ is an intrinsic gauge parameter unique to a superconductor.
Since $J_1 >J_2$ in Fig.~\ref{fig:tdepj},
we define the phase across the junction in terms of $\varphi_{1 L(R)}$ by
\begin{align}
\delta\varphi\equiv \varphi_{1L}-\varphi_{1R}.
\end{align}
The phase difference in the second band is described by
\begin{align}
\varphi_{2L}-\varphi_{2R}=& \delta \varphi - 2\delta\theta.
\end{align}
The integer numbers $n_L$ and $n_R$ in Eqs.~(\ref{trs_l}) and (\ref{trs_r})
are embedded into $\theta_L$ and $\theta_R$,
respectively.
The Josephson current becomes
\begin{align}
J=&J_1 \sin\delta\varphi
+ J_2 \sin(\delta\varphi- 2 \delta\theta) +J_{12} \sin(\delta\varphi - \delta\theta),\\
=&J_1 \sin\delta\varphi
+ J_2\left[ \sin\delta\varphi\cos (2 \delta\theta) - \cos\delta\varphi \sin (2\delta\theta)
\right] \nonumber\\
&+J_{12}\left[ \sin\delta\varphi\cos \delta\theta - \cos\delta\varphi \sin \delta\theta
\right].\label{cpr}
\end{align}
In what follows, we discuss characteristic properties of
current-phase relationship (CPR) for several examples of the Josephson
junctions.

\textbf{(i)} $s_{++}/s_{++}$\\
 A two-band superconducting state with $\varphi_1=\varphi_2$ is called an
 $s_{++}$ state in recent literature and is described by $\theta=0$ or $\pi$ in this paper.
We first consider a junction consisting of two $s_{++}$
superconductors with $\delta\theta=0$. The resulting current given by
\begin{align}
J=(J_1+J_2+J_{12}) \sin\delta\varphi, \label{cpp1}
\end{align}
is sinusoidal as a function of the phase difference.
It is possible to consider another $s_{++}/s_{++}$ junction by choosing
$\theta_L=\pi$ and $\theta_R=0$. The current in such
case results in
\begin{align}
J=(J_1+J_2-J_{12}) \sin\delta\varphi. \label{cpp2}
\end{align}
Although the CPR is sinusoidal, the last term changes its sign.
In an isolated superconductor, the gauge parameter $\theta$ does not affects any
observable values because it can be removed by a gauge transformation.
In a Josephson junction, however, the relative difference in $\theta$ in the two
superconductors $\delta\theta$ modifies the Josephson current.
In other words, any spatially uniform
gauge transformations cannot remove $\delta\theta$.

 \textbf{(ii)} $s_{+-}/s_{+-}$\\
A two-band superconducting state with $\varphi_1=\varphi_2+\pi$ is called
an $s_{+-}$ state and is described by $\theta=\pm \pi/2$.
In a junction consisting of two $s_{+-}$ superconductors with $\theta_L=\theta_R=\pi/2$,
the CPR is given by Eq.~(\ref{cpp1}).
However, when we choose $\theta_L=\pi/2$ and $\theta_R=-\pi/2$,
the CPR becomes Eq.~(\ref{cpp2}).
As well as in the case of (i), CPR is sinusoidal in both cases.

\textbf{(iii)} $s_{+-}/s_{++}$

Finally, we consider a junction consisting of an $s_{++}$ superconductor ond
an $s_{+-}$ superconductor.
The results are summarized as
\begin{align}
J=(J_1-J_2) \sin\delta\varphi \mp J_{12}\cos\delta\varphi,
\end{align}
for $\theta_L=\pm\pi/2$ and $\theta_R=0$.
 When we choose $\theta_L=\pm\pi/2$ and $\theta_R=\pi$, the CPR becomes
\begin{align}
J=(J_1-J_2) \sin\delta\varphi \pm J_{12}\cos\delta\varphi.
\end{align}
The sign change of $J_2$ means that the $\pi$ phase difference in the pair potentials
in the second band.
The most important feature is the appearance of $\cos\delta\varphi$ term, which
suggests the breakdown of time-reversal
symmetry even at $\delta\varphi=0$.
According to Eq.~(\ref{cpr}), $\cos\delta\varphi$ term always appears when
$\delta\theta$ is neither 0 nor $\pi$.
Namely, the Josephson current flows even at $\delta\varphi=0$.
In such case, the all the phase factors in $H_{\mathrm{JJ}}$ in Eq.~(\ref{h_jj})
cannot be removed by a uniform gauge transformation.
Thus $\cos\delta\varphi$ term in the Josephson current is a result of
time-reversal symmetry breaking in $H_{\mathrm{JJ}}$.

%
\begin{figure}[tbh]
\begin{center}
\includegraphics[width=7.5cm]{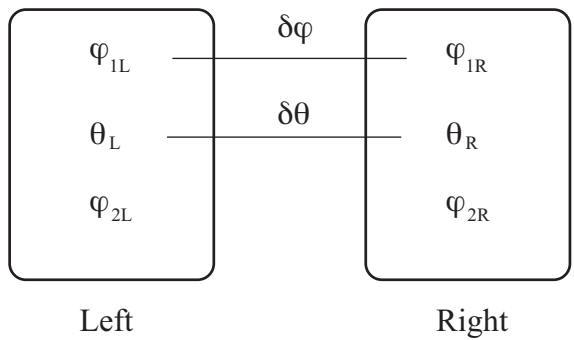}
\end{center}
\caption{
In a superconductor, the phase difference between the two pair potential
 $\varphi_{1j}-\varphi_{2j}$ is related to $\theta_j$ in
 Eqs.~(\ref{trs_l}) and (\ref{trs_r}) which
hold true in the presence of time-reversal symmetry.
The Josephson current is represented as a function of $\delta\varphi$ and $\delta\theta$.
 }
\label{fig:phase}
\end{figure}

\section{Junction energy}

The Josephson current is relating to the junction energy $E_{\mathrm{JJ}}$ by
$ J(\delta\varphi)= e \partial_{\delta\varphi} E_{\mathrm{JJ}}$.
Namely, the Josephson current vanishes at the ground state.
From the relation, the junction energy can be described by
\begin{align}
e E_{\mathrm{JJ}} =&
-J_1 \cos\delta\varphi - J_2 \cos(\delta\varphi - 2\delta\theta) \nonumber\\
&- J_{12} \cos(\delta\varphi - \delta\theta),\\
\geq &-(J_1+ J_2 +|J_{12}| ).
\end{align}
As shown in Eqs.~(\ref{j1})-(\ref{j12}), $J_1$ and $J_2$ are always positive but $J_{12}$ can change its sign.
We first discuss the case of $J_{12}>0$.
As shown in Fig.~\ref{fig:tdepj}, the first term proportional to $J_1$
is the most dominant in our model and takes its minimum at $\delta\varphi=0$.
The second and the third terms, however, take their minima at $\delta\varphi=2\delta\theta$
and $\delta\varphi=\delta\theta$, respectively.
In this paper, we fix $\delta\theta$ at a certain value and calculate the Josephson current.
As we have discussed in Sec.~\ref{model}, the gauge parameter $\theta$
does not affect physics of a single superconductor
in the presence of time-reversal symmetry because it is
possible to remove it by a unitary transformation.
In addition, $\theta$ originates from the difference in phases of two atomic
orbital functions~\cite{leggett:PTP1966,asano:prb2018}.
Therefore, it is reasonable to consider that two superconductors
would adjust their gauge parameters $\delta\theta$
to minimize the junction energy spontaneously.
Indeed, $E_{\mathrm{JJ}}$ is at its absolute minimum
at $(\delta\varphi, \delta\theta)=(0,0)$ for $J_{12}>0$.

For $J_{12}<0$, the junction energy $E_{JJ}$
takse its minimum at $(\delta\varphi, \delta\theta)= (0,\pi)$.
Namely, the sign change of $J_{12}$ is absorbed as $\pi$ shift of $\delta\theta$.
The CPR remains sinusoidal even at such a nontrivial solution because $\pi$ shift in $\theta$ does
not break time-reversal symmetry as shown in Eq.~(\ref{trs_cond}).
Thus we conclude that the CPR of a Josephson junction consisting of two
time-reversal two-band superconductors is sinusoidal as shown in Eq.~(\ref{cpp1})
when the two superconductors minimumize the junction energy spontaneousely by
adjusting thier gauge parameter $\delta\theta$.

\section{Conclusion}
We studied Josephson effect in a tunnel junction consisting of two
time-reversal two-band superconductors theoretically.
We introduce the equal-time spin-singlet $s$-wave pair potentials in each
conduction band, the band hybridization as well as the band asymmetry.
On the basis of a standard current formula, the Josephson current
is described by the anomalous Green's functions on either side of the junction.
The Josephson current depends not only on the phase difference across the junction
but also on the difference of the gauge parameter in the two superconductors.
The current-phase relationship deviates from the sinusoidal function
when the two gauge parameters are not equal to each other.
The dependence of the Josephson current on temperature,
the behavior of the gauge parameters in real junctions, and effects of odd-frequency Cooper pairs
on the Josephson current are also discussed.

\begin{acknowledgments}
The authors are grateful to R. Sitro, C. Noce, and A. Romano for useful discussions.
This work was supported by Topological Materials Science (Nos.~JP15H05852 and JP15K21717)
from the Ministry of Education, Culture, Sports, Science and Technology (MEXT) of
Japan, JSPS Core-to-Core Program (A. Advanced Research Networks),
JSPS and Russian Foundation for Basic Research under Japan-Russia Research Cooperative Program.
\end{acknowledgments}



%


\end{document}